# Gamma Ray Counters to Monitor Radioactive Waste Packages in the MICADO Project

Luigi Cosentino[1], Martina Giuffrida[1,2], Sergio Lo Meo[3], Fabio Longhitano[1], Alfio Pappalardo[1,§], Giuseppe Passaro[1], and Paolo Finocchiaro[1,2,*]

1   INFN Laboratori Nazionali del Sud, 95123 Catania, Italy; cosentino@lns.infn.it; giuffrida@lns.infn.it; fabio.longhitano@lns.infn.it; alfio.pappalardo@eli-np.ro; passaro@lns.infn.it; finocchiaro@lns.infn.it
2   Centro Siciliano di Fisica Nucleare e Struttura della Materia, 95123 Catania, Italy
3   ENEA Centro Ricerche, 40129 Bologna, Italy; sergio.lomeo@enea.it
§   now at ELI-NP, Magurele, Rumania
*   Correspondence: finocchiaro@lns.infn.it

**Abstract:** One of the goals of the MICADO Euratom project is to monitor the gamma rays coming out of radioactive waste drums in storage sites on a medium/long term basis. For this purpose 36 low-cost gamma ray counters were designed and built to act as a demonstrator. These counters, named SciFi, are based on a scintillating fiber readout at each end by a silicon photomultiplier, assembled in a robust arrangement in form of 80 cm long pipes. Several counters will be placed around radwaste packages in order to monitor the gamma dose-rate by collecting a continuous data stream. The 36 sensors were thoroughly tested with a $^{22}$Na, a $^{137}$Cs and an AmBe sources, the results are quite satisfactory and the next step will be the test in a real environment.

## 1. Introduction

The MICADO (Measurement and Instrumentation for Cleaning And Decommissioning Operations) Euratom (EU) project is aimed at the full digitization of low- and intermediate-level radioactive waste management [1,2]. Following a complete active and passive characterization of the radwaste drums with neutrons and gamma rays, the project contemplates a longer-term monitoring phase in the Work Package 7 by means of low-cost dedicated detectors for neutrons (named SiLiF [3]) and for gamma rays (named SciFi). A continuous automatic monitoring of the radwaste drums after their characterization represents an added value in terms of safety and security, and the availability of continuous streams of counting-rate data around each drum would be a comfortable tool toward the transparency, which now more than ever is a relevant topic of the nuclear industry with respect to the common people environment-aware [1,3-7].

The radiological monitoring of radwaste has to be based on the measurement of gamma rays and neutrons, because they are penetrating and thus more easily detectable out of the drums. Due to the foreseen mass deployment, the sensors have to be reasonably low-cost and configurable in a modular and scalable fashion, so that one can tailor the system to small, medium and large scale storage configurations. The proposed monitoring system is based on detectors which can be easily installed and/or reassembled in different geometrical configurations, as they are mechanically very simple and are based on commercial electronics.

The neutron monitoring detectors have already been described in reference [3], and in this paper we describe the SciFi monitoring detectors devoted to the gamma radiation: following a description of their operational principles and mechanical setup, some simulation results are described followed by the results of the characterization and a few tests of 36 SciFi detector units. Finally, an example is provided with the evaluation of the expected performance and sensitivity in a realistic case.

## 2. Materials and Methods

When facing the development of gamma ray monitoring detectors for a possible mass deployment around radwaste drums, we made the following considerations:

- the sensors should be simple and robust;
- they do not need to have spectroscopic features but can be simple counters;
- they can have low intrinsic efficiency, as they can measure for long time spans and be sensitive enough;
- low efficiency also implies the capability to stand in high radiation fluxes without being saturated;
- they should possibly be spatially extended to cover a wide region of a drum;
- they should likely be based on commercially available technology;
- they should be reasonably inexpensive.

In light of all this we opted for a solution based on a length of plastic scintillating fiber read out at both ends by Silicon PhotoMultipliers (SiPM), which represents a low-cost solution based on commercial products. Moreover, such a solution has a high degree of modularity in terms of length, shape and number of detectors.

### 2.1. The scintillating fiber

The chosen scintillating fiber, whose main characteristics are listed in **Table 1**, is the BCF-20 produced by Saint Gobain [8]. We tested both 1 mm and 3 mm diameter fibers, and chose the latter which has a higher detection efficiency as the average energy deposited by the detected gamma rays is larger. Moreover, even though the 3 mm fiber is stiffer than the 1 mm one, it is still flexible enough for a possible future solution in a curved or round shape.

**Table 1.** Characteristics of the chosen scintillating fiber as reported by the manufacturer.

| Fiber Type | Emission Peak, nm | Decay Time, ns | Light yield (#Photons per MeV) | Trapping efficiency | Attenuation length |
|---|---|---|---|---|---|
| BCF-20 | 492 | 2.7 | ~8000 | ≈ 6% on each side | 3.3 m |

The operating principle is the following: whenever a gamma ray interacts with the fiber it deposits a variable amount of kinetic energy which gives rise to a short flash of scintillation light. A fraction of these scintillation photons is trapped into the fiber and propagates toward both ends. If the number of these photons is large enough to produce a signal above a predefined threshold simultaneously on both fiber ends, such a coincidence event is considered as the detection of a gamma ray and can be counted. By using the GEANT4 code [9] we simulated the interaction of gamma rays of a few selected energies with fibers of 1 and 3 mm diameter, recording the energy deposited on the fiber event by event whenever the deposition was at least 50 keV. Indeed, below 50 keV the average number of photons detected on the SiPM would be < 4, as will be explained in sections 2.2 and 2.3. Then, for each impinging gamma energy we calculated the average deposited energy, and the results are plotted in **Figure** 1. Apart from very low energy cases where the photoelectric effect could take place, the interaction occurs through compton effect and the energy deposition is done by the scattered electron. The plateaux in **Figure** 1 are due to the electron escaping the fiber, and reflect the different fiber diameters. The same simulation allowed us to calculate the expected intrinsic detection efficiency as the ratio between the reported counts and the number of impinging gamma rays. The resulting plots, reported in **Figure** 2, fully justify the choice of the 3 mm diameter fiber for the SciFi detectors.

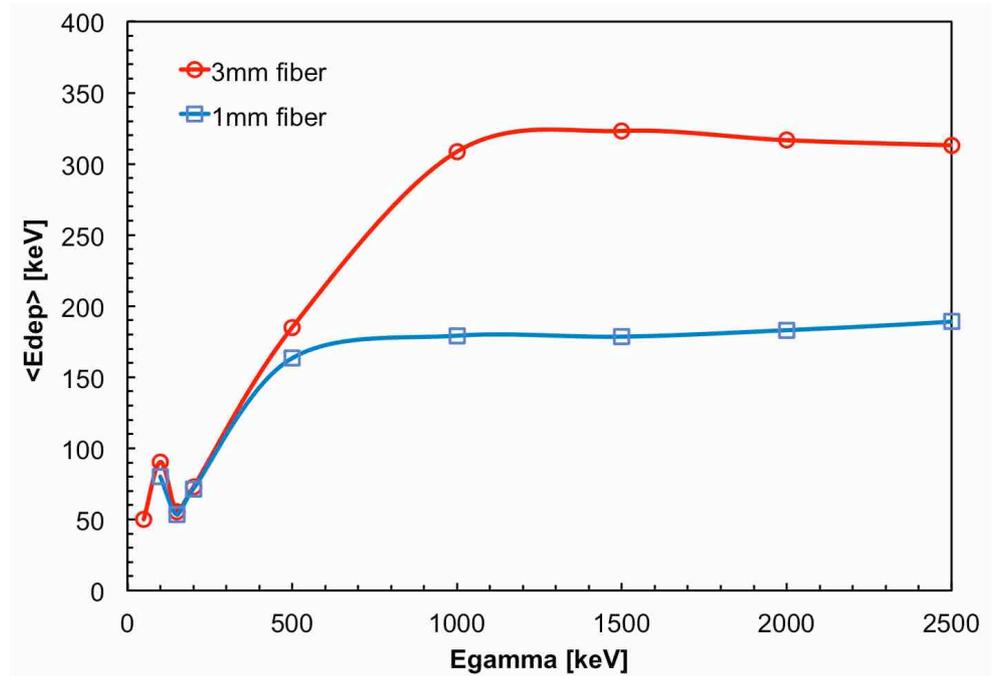

**Figure 1.** Simulation of the average energy deposited in a 1 mm and a 3 mm fiber as a function of the impinging gamma ray energy (threshold at 50 keV). The lines are spline interpolations.

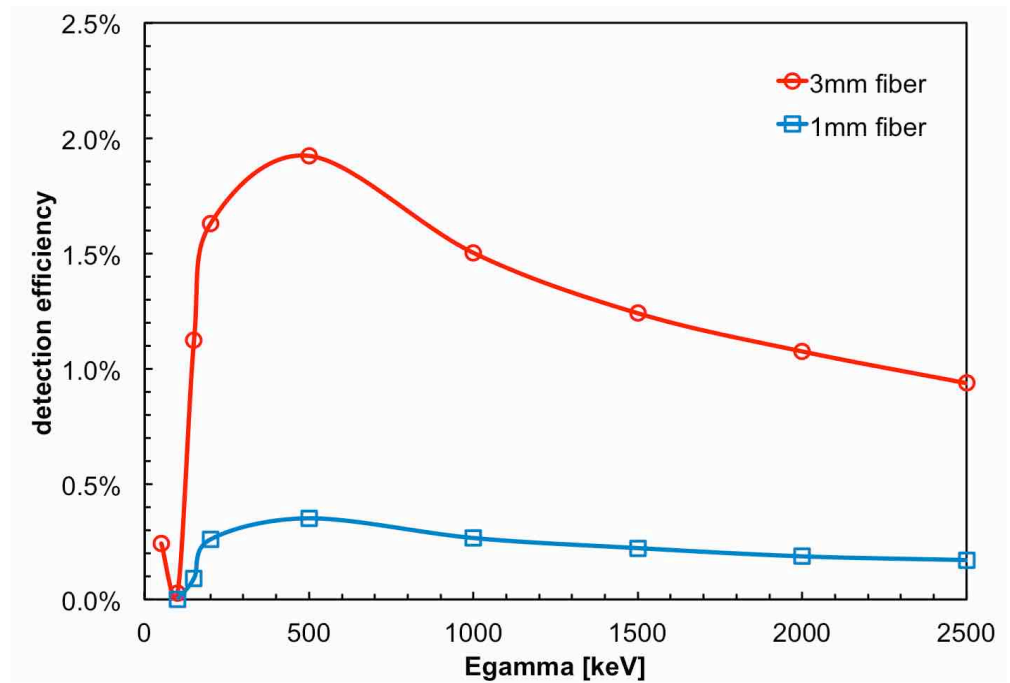

**Figure 2.** Simulation of the intrinsic detection efficiency of a 1 mm and a 3 mm fiber for several values of the impinging gamma ray energy. The lines are spline interpolations.

### 2.2. The SiPM photodetector

The SiPM photodetector, capable of counting single photons, is an intrinsically noisy device whose noise level can be dramatically reduced by enforcing a suitable threshold on its output signals [10-20]. The chosen SiPM is the MicroFC-30035-SMT, produced by ON Semiconductor [21], whose main characteristics are listed in **Table 2**. For all the tests described in this work we made use of a home made voltage amplifier, which features a 200x gain and a 4 GHz bandwidth. The SiPM bias, according to the manufacturer's specifications, was set at a 2.5 V overvoltage i.e. 27 V bias.

**Table 2.** Main characteristics of the MicroFC-30035-SMT SiPM.

| Sensor size | Microcell size | Number of microcells | Breakdown voltage | Overvoltage | PDE at 420 nm | PDE at 500nm |
|---|---|---|---|---|---|---|
| 3x3 mm$^2$ | 35 $\mu$m | 4774 | 24.5 V | 2.5 V | 30% | 20% |

In order to choose an appropriate threshold level for the SiPM we studied the noise counting rate as a function of the threshold, and the resulting typical staircase plot is reported in **Figure** 3. Each step corresponds to one detected photon, and we decided to use a threshold at 175 mV that corresponds to ≈ 3.5 photons and reduces the noise rate by three orders of magnitude down to about 350 counts per second. As the signals have a duration of about 30 ns, a duration of 100 ns for the coincidence window between the two SiPMs makes the probability of spurious coincidences negligible.

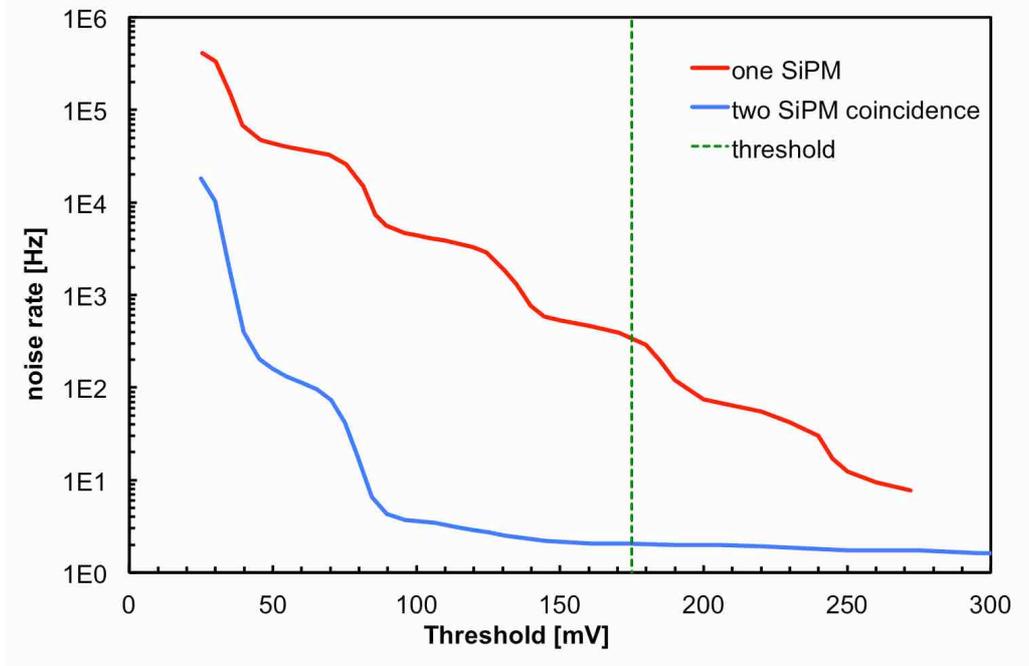

**Figure 3.** Red line: Typical SiPM staircase plot, obtained by reporting the noise counting rate as a function of the threshold. The final chosen threshold was 175 mV, i.e. 4 photons. Blue line: same plot with the SiPMs coincidence signal from a SciFi detector. The constant background above the threshold is physical from cosmic rays and environmental radiation.

### 2.3. The SciFi detector assembly

In order to decide the fiber length we calculated the probability of detection as a function of the impact position, of the deposited energy and of the threshold selected on the SiPMs, taking into account the scintillation light yield, the light trapping efficiency, the fiber attenuation length, and the photon detection efficiency (PDE) of the SiPM. Using input values taken from **Table 1** and **Table 2**, and assuming the threshold at ≥ 4 photons on the SiPMs, we produced the corresponding plots for fiber length of 80, 120 and 300 cm, in two cases of 100 and 200 keV energy deposited in the fiber. The plots are reported respectively in **Figure 4**, **Figure 5** and **Figure 6**. One can immediately see that with 200 keV deposited energy the response is uniform for the 80 and 120 cm fibers, whereas it is acceptable for the 300 cm one. Conversely, with 100 keV deposited energy the 80 and 120 cm fibers respond almost uniformly, even though losing 15-20% efficiency; the 300 cm fiber in such a case is barely acceptable.

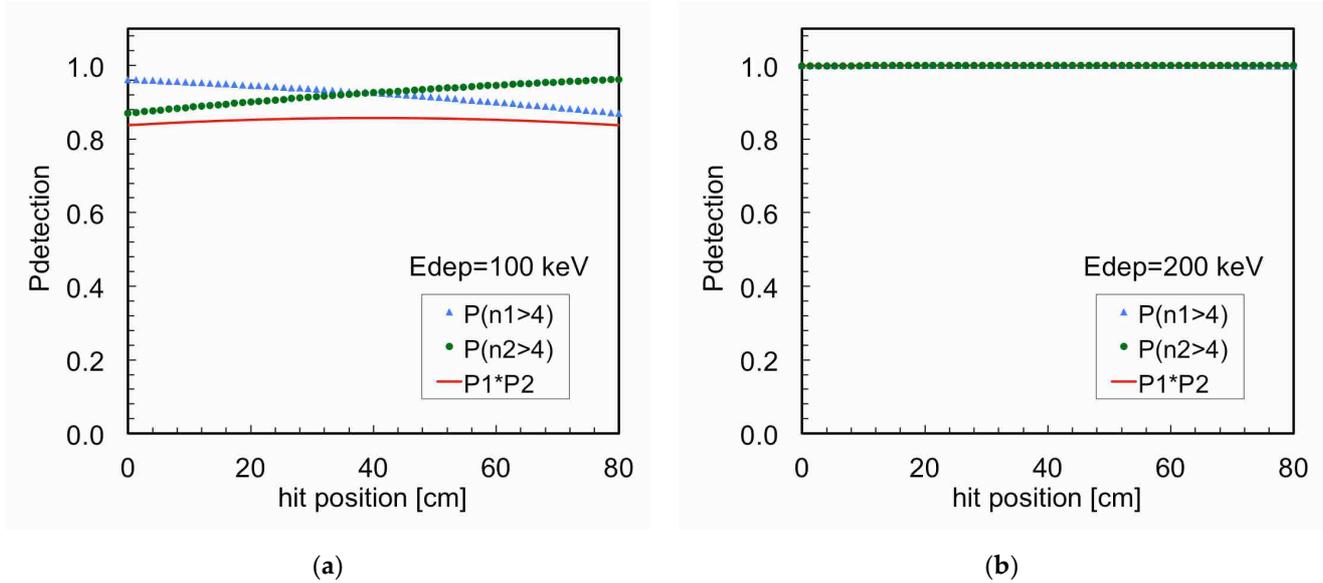

(a)                                                          (b)

**Figure 4.** Detection probability of a gamma ray as a function of the impact position on a 80 cm long fiber, calculated as the product of the probabilities of getting at least 4 photons detected on each SiPM. **(a)** When the gamma ray deposits 100 keV. **(b)** When the gamma ray deposits 200 keV.

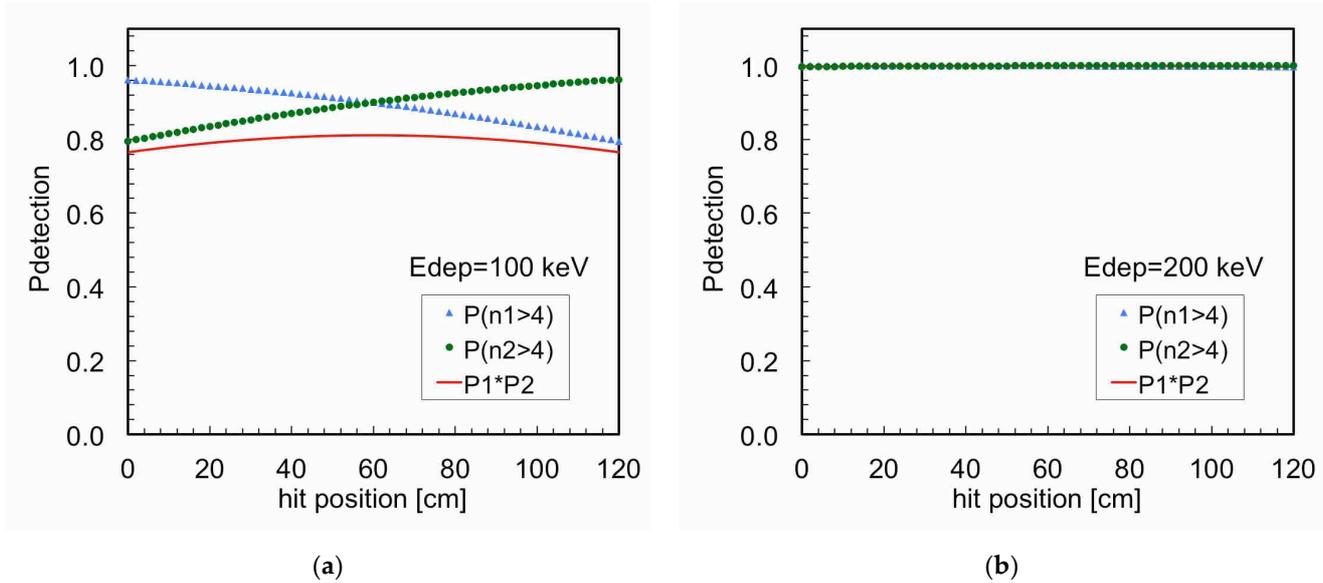

(a)                                                          (b)

**Figure 5.** Detection probability of a gamma ray as a function of the impact position on a 120 cm long fiber, calculated as the product of the probabilities of getting at least 4 photons detected on each SiPM. **(a)** When the gamma ray deposits 100 keV. **(b)** When the gamma ray deposits 200 keV.

We opted for the 80 cm length, even in light of the height of the standard 120 and 210 liters radwaste drums (76 and 88 cm). The scintillating fiber was allocated inside a 2 cm diameter and 1 mm thick aluminum pipe, and held in place by two cylindrical holders designed to host a circular PCB with the SiPM and its support circuitry. The holder has a central hole to allow for the fiber-to-SiPM alignment and optical coupling by means of a tiny grease drop, and two side grooves to allow for the passage of cables (for this prototyping phase we chose the grease for reversibility, an optical glue will be the final solution). Two light-tight rubber caps complete the setup, with three cables coming out of one single side for the common voltage bias and the two output signals. A sketch and three pictures of the SciFi detector components are shown in **Figure 7**. We remark that the attenuation of gamma rays when crossing the aluminum pipe is a few percent at very low energy and goes down to about 1-2% at higher energy (**Figure** 8) [22].

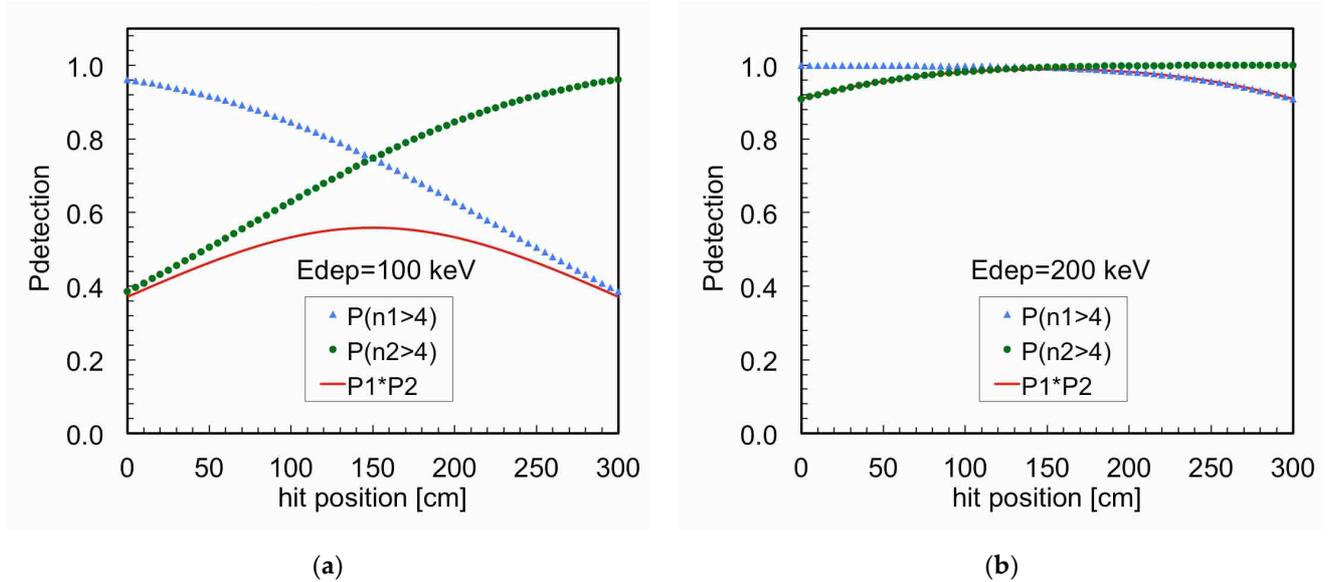

(**a**)                                                            (**b**)

**Figure 6.** Detection probability of a gamma ray as a function of the impact position on a 300 cm long fiber, calculated as the product of the probabilities of getting at least 4 photons detected on each SiPM. **(a)** When the gamma ray deposits 100 keV. **(b)** When the gamma ray deposits 200 keV.

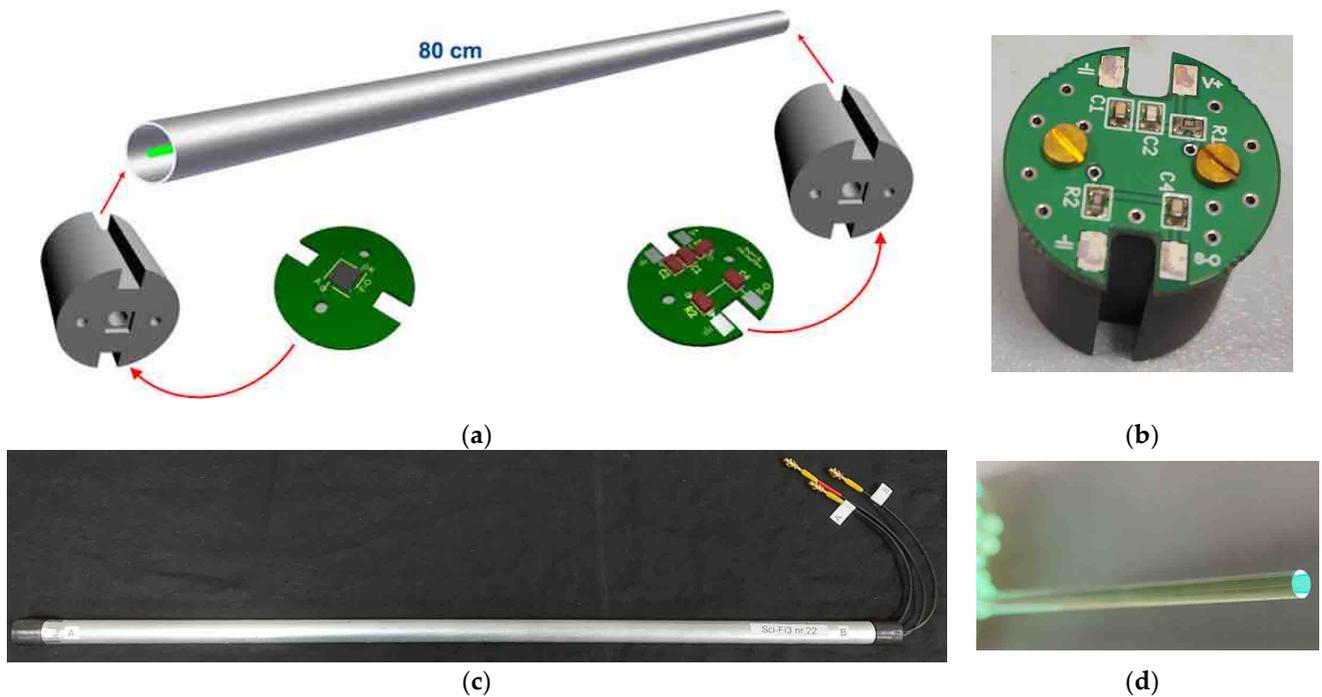

(**a**)                                                            (**b**)

(**c**)                                                            (**d**)

**Figure 7. (a)** Sketch of the components of the SciFi detector assembly. **(b)** A fiber holder with the SiPM PCB in place. **(c)** A detector fully assembled. **(d)** A 3 mm diameter fiber.

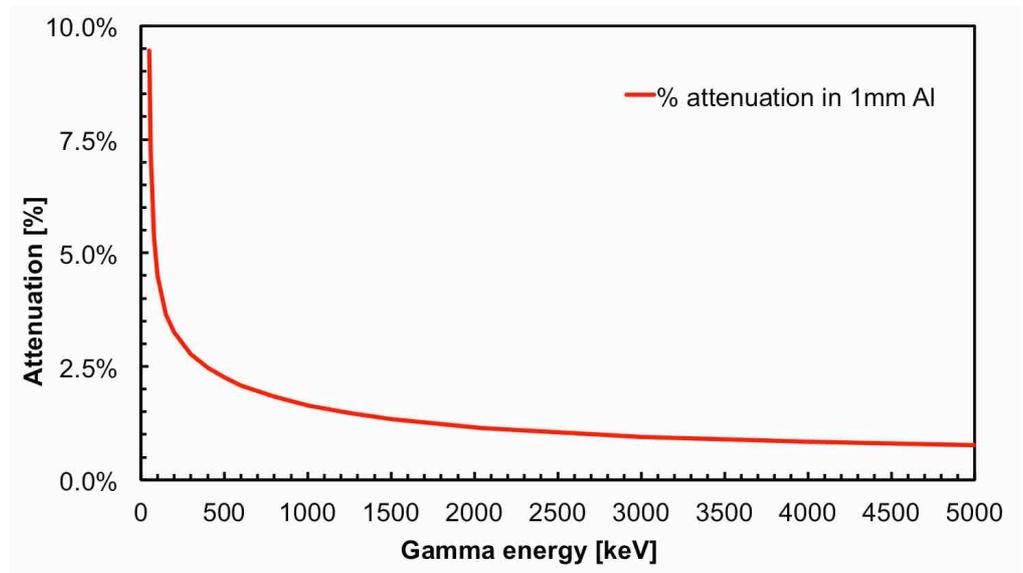

**Figure 8.** Attenuation of gamma rays in 1 mm aluminum as a function of the energy [22].

## 3. Results

Thirty-six SciFi detectors were assembled as shown in the previous section, and in order to characterize them we made several tests with standard laboratory gamma sources and a high activity gamma and neutron source. The electronic setup, sketched in **Figure 9**, was quite simple: the outputs of the SiPM amplifiers were connected to two discriminators, with threshold equal to 175 mV. The two logic outputs were used as inputs to a coincidence unit with a 100 ns window, and the final output was sent to a counter.

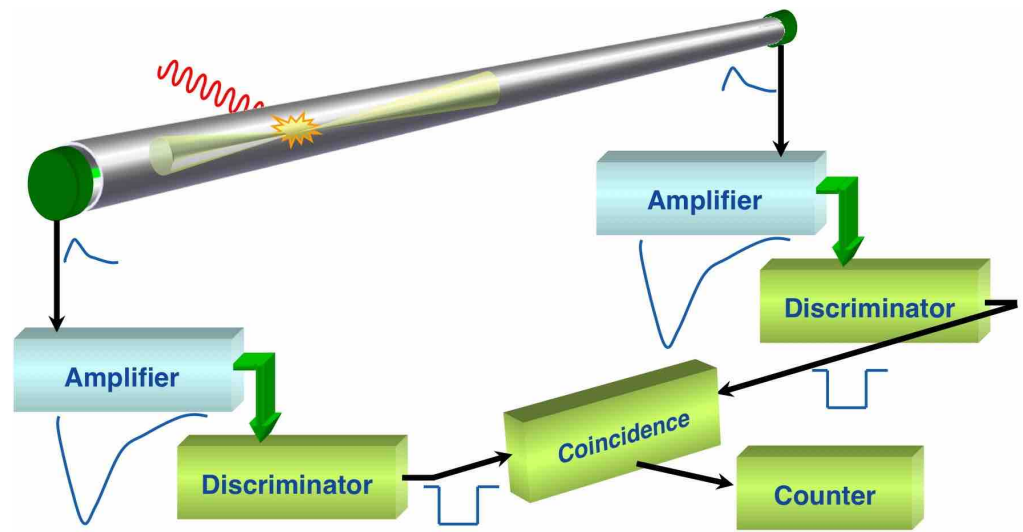

**Figure 9.** Sketch of the electronic setup for the SciFi detector.

### 3.1. Pointlike source at short distance

A first test to validate the detector behavior was done by means of a pointlike $^{22}$Na source of activity A ≈ 42 kBq, by counting the detected gamma rays as a function of the distance in the range of few centimeters. The source was placed at several distances from the fiber, and the number of counts was recorded. The source emits gamma rays of 1274 keV (99.9% branching ratio) and also β+ particles (90.3% branching ratio) which immediately annihilate giving rise to a pair of 511 keV gamma rays. When the distance r between source and fiber is much smaller than the fiber length, this one can be assumed as infinite and the geometrical efficiency ε can be calculated as

$$\varepsilon \approx w/2\pi r, \tag{1}$$

where w is the fiber diameter. The expected count rate C can be calculated, for instance at 1 cm, by means of the activity A and the simulated detection efficiency $\varepsilon_{det}$ at 511 and 1274 keV:

$$C(1\ cm) \approx A \times \varepsilon(1\ cm) \times [2 \times 0.903 \times \varepsilon_{det}(511\ keV) + 0.999 \times \varepsilon_{det}(1274\ keV)] \approx 100\ cps \tag{2}$$

that is what we obtained and corresponds to an equivalent dose of about 118 μSv/h [23]. A 1/r fit reproduces perfectly the measured count rates, as shown in **Figure** 10.

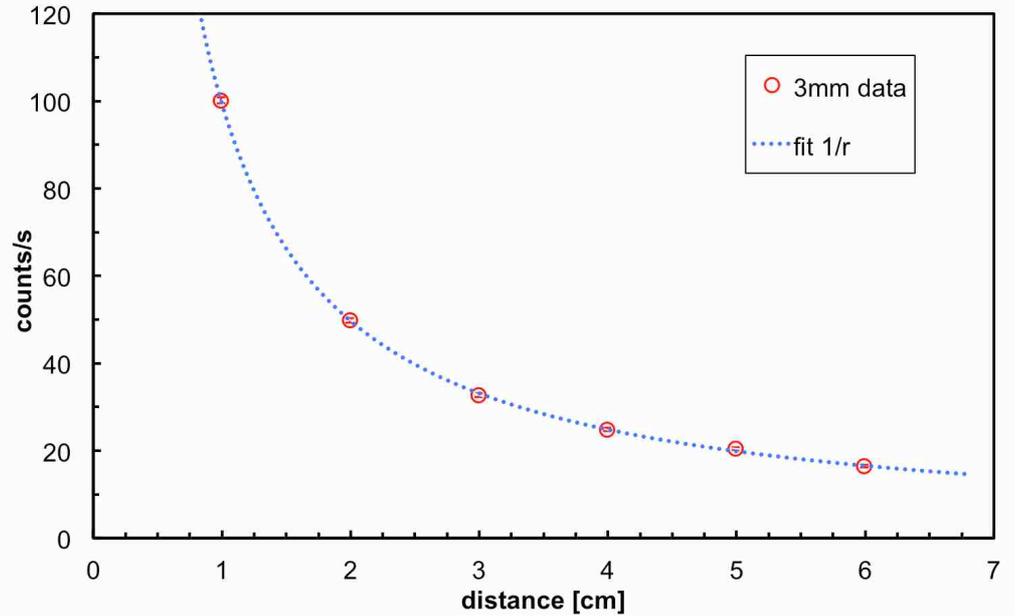

**Figure** 10. The observed count rate as function of the distance between the pointlike $^{22}$Na source and the 3 mm diameter fiber. Also shown is a 1/r fit.

### 3.2. Detection efficiency measurement

For a better check of the simulation results, and in order to verify the uniformity of the response within the set of SciFi detectors, we assembled a simple setup with a point-like $^{137}$Cs gamma source, with activity A = 1.49 Mbq, and a detector holder that kept the midpoint of the fiber at distance d = 52 cm from the source (**Figure** 11). The counting rate of each detector was measured in 200 s, as well as the background rate and the signal-to-background ratio, and the results are plotted in **Figure** 12. The nonuniformity of the response is due to tiny differences between the individual SiPMs, which are amplified by their exponential multiplication mechanism, and slight imperfections of the optical coupling. However, the nonuniformity is quite limited and reasonable. Notice that the background count level is the same as shown by the blue line in **Figure** 3, as it represents the physical background due to cosmic rays and natural radioactivity.

The net measured counting rates, i.e. after subtracting the background, were used to calculate the detection efficiency with gamma rays of 662 keV energy. The geometrical efficiency ε in this configuration can be calculated starting from a barrel-like solid angle covering (**Figure** 13), subtracting the fraction covered by the two spherical caps, and then dividing by the number of fibers that would cover the entire cylinder. The relevant quantities are d = 52 cm, fiber length f = 80 cm, fiber diameter c = 0.3 cm, r = 65.6 cm, h = 25.6.

$$\Omega_{cap} / \Omega_{total} = 2\pi rh/4\pi r^2 = h/2r \approx 0.195, \tag{3}$$

$$\Omega_{cylinder} / \Omega_{total} \approx 1 - 2 \times \Omega_{cap} / \Omega_{total} \approx 0.610, \tag{4}$$

$$N_{fibers} \approx 2\pi r/c \approx 1089, \tag{5}$$

$$\varepsilon = (\Omega_{cylinder} / \Omega_{total}) / N_{fibers} \approx 0.00056, \tag{6}$$

The decay branching ratio of $^{137}$Cs into 662 keV gamma rays is 85%, and the attenuation in the aluminum pipe is $\approx 2.0\%$, therefore the expected number of impinging gamma rays on the fiber per second was

$$N_{gamma} \approx A \times 0.85 \times (1-0.02) \times \varepsilon \approx 694 \tag{7}$$

The average net counting rate from the 36 fibers was 11.43 counts per second (cps), therefore the average detection efficiency resulted

$$\varepsilon_{det}(662 \text{ keV}) \approx 11.43 / 694 \approx 1.65\%, \tag{8}$$

in reasonable agreement with the 1.8% value estimated from the simulation. The detection efficiency for all the fibers is reported in **Figure** 14.

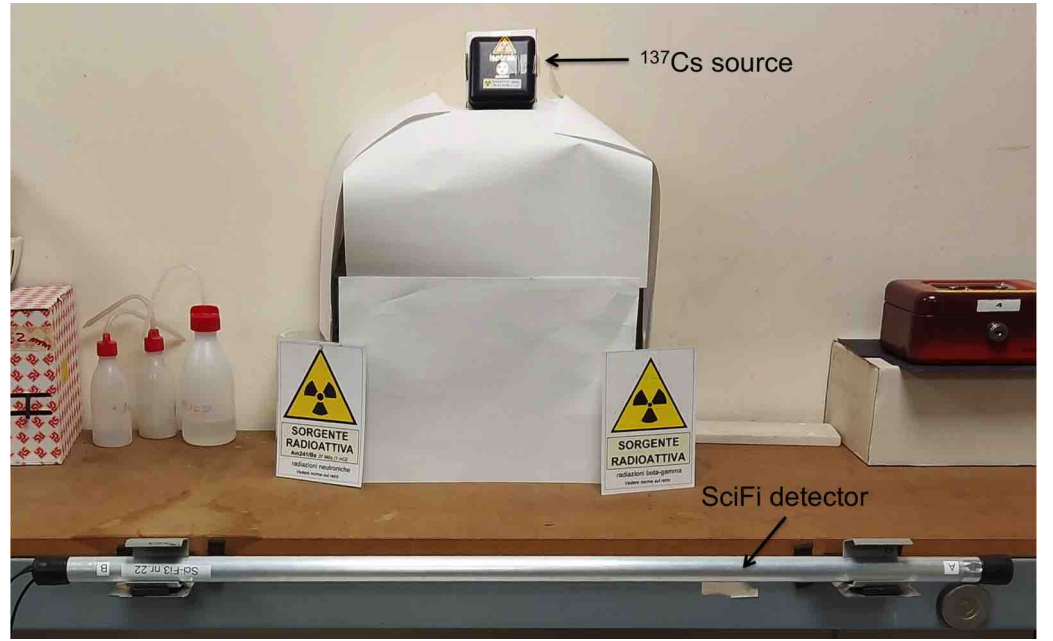

**Figure** 11. Measurement setup with a pointlike 1.49 Mbq $^{137}$Cs gamma source and a detector holder. The distance between the source and the midpoint of the fiber was d = 52 cm.

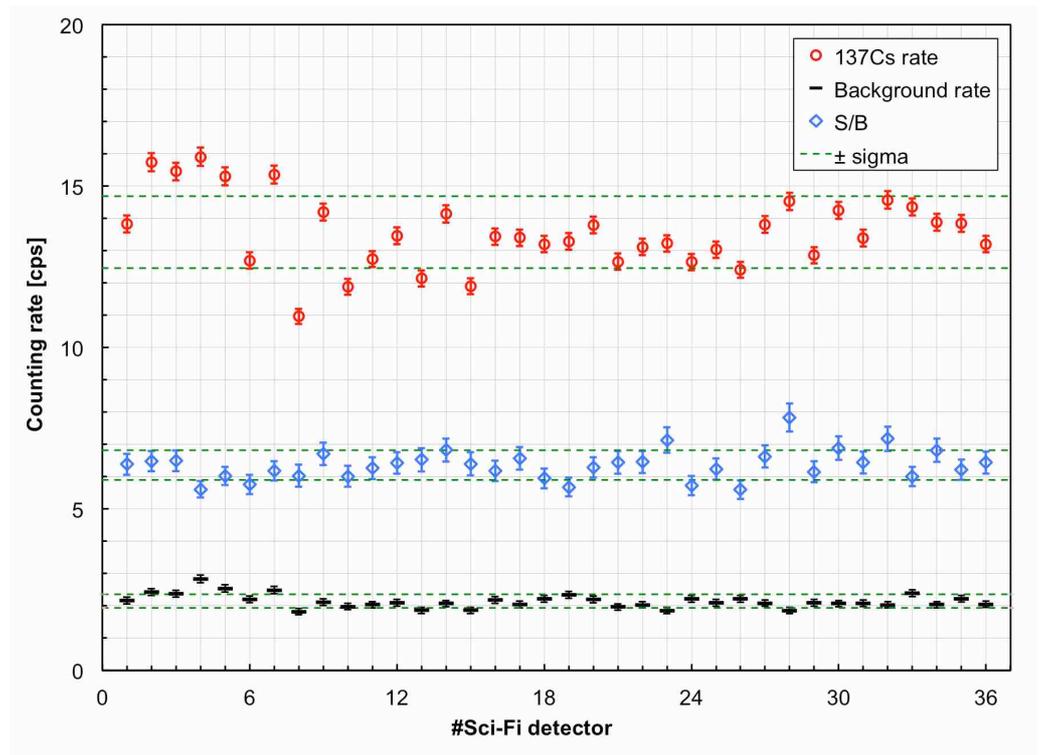

**Figure** 12. The counting rate of the 36 SciFi detectors, measured with the ¹³⁷Cs source in 200 s in the setup of **Figure** 11, along with the background rate and the signal-to-background ratio. The error bars represent the statistical uncertainty, the green dashed lines indicate ±1 standard deviations from the average.

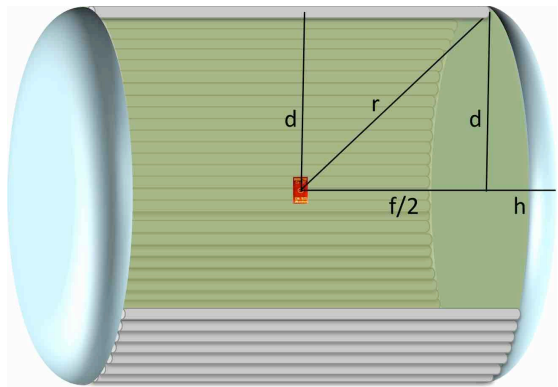

**Figure** 13. Sketch for the calculation of the geometrical efficiency of the SciFi detector in the setup of **Figure** 11.

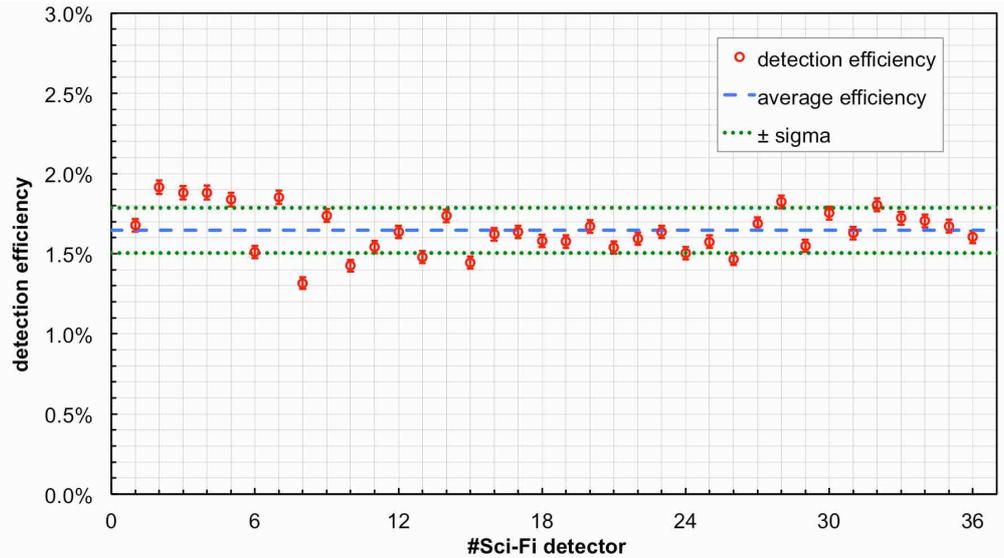

**Figure 14.** Measured detection efficiency with 662 keV gamma rays from the $^{137}$Cs source. The error bars represent the statistical uncertainty; the blue dashed line is the average efficiency; the green dotted lines indicate ±1 standard deviations from the average.

### 3.3. Tests in a more realistic setup

A realistic setup resembling the emission from a radwaste drum was employed for a series of tests of the SciFi detectors. We made use of an intense AmBe neutron source, installed in an experimental hall at INFN Laboratori Nazionali del Sud (LNS), which emits $2.2 \times 10^6$ neutrons/s by exploiting alpha particles from the $^{241}$Am decay to induce the $^9$Be($\alpha$,n) reaction. In order to produce such an amount of neutrons one needs a considerable quantity of $^{241}$Am, which is highly radioactive as it emits 59 keV gamma rays. Indeed the activity of our source is 34 GBq, and it is enclosed in a 95x75x85 cm³ iron box along with its moderator (**Figure** 15). The source is surrounded by a first polyethylene case followed by 30 cm thick paraffin which slow down the high energy neutrons whose initial kinetic energy extends up to 10 MeV. The outer 5 cm of the shielding are made from borated paraffin that absorbs the vast majority of the outgoing thermalized neutrons. The energy spectrum of the gamma rays coming out of the box has several main components:

- 59 keV from the decay of $^{241}$Am;
- 478 keV from neutron capture in boron (component of the borated paraffin);
- 511 keV from e+e- pair production followed by positron annihilation;
- 2200 keV from neutron capture in hydrogen (component of paraffin and polyethylene);
- 4438 keV from the $^9$Be($\alpha$,n)$^{12}$C*;
- a continuum due to the compton scattering in the source assembly materials.

All the SciFi detectors were tested on top of the source box, and the measured counting rate, the background and the signal-to-background ratio are reported in **Figure** 16. For each detector we calculated the ratio between the observed counting rates respectively with the $^{137}$Cs and the AmBe sources. The results, plotted in **Figure** 17, indicate a quite reasonably constant behavior of this ratio.

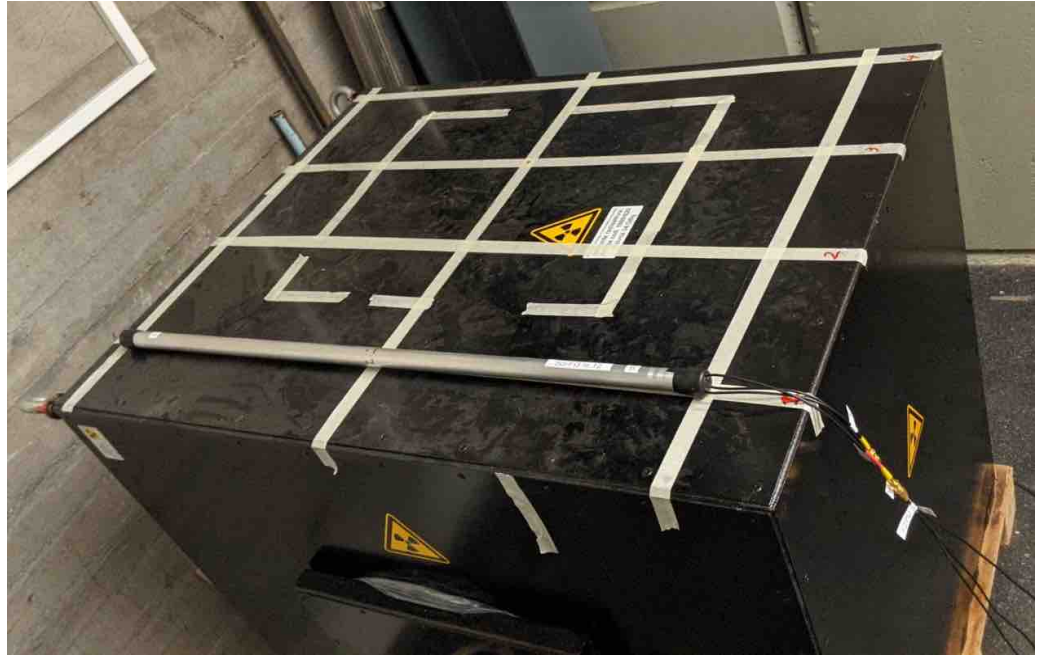

**Figure 15.** The AmBe neutron and gamma source enclosed in a 95x75x85 cm³ iron box along with its moderator. A SciFi detector in one of the measurement positions is also shown.

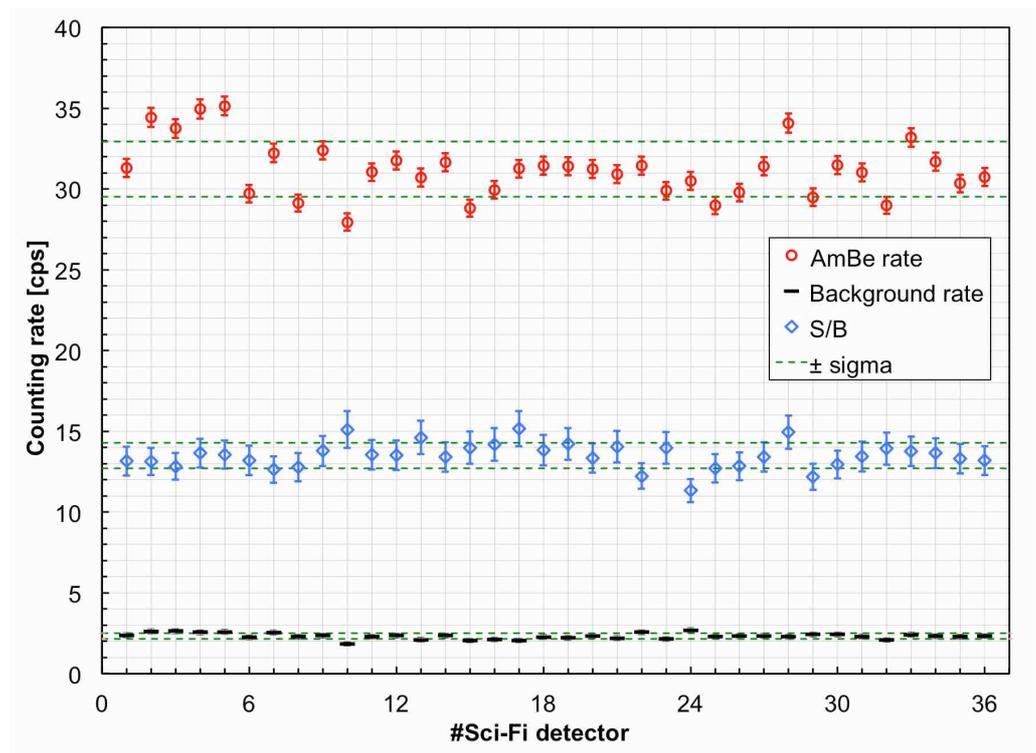

**Figure 16.** The counting rate of the 36 SciFi detectors, measured on top of the AmBe source in 100 s, along with the background rate and the signal-to-background ratio. The error bars represent the statistical uncertainty, the green dashed lines indicate ±1 standard deviations from the average.

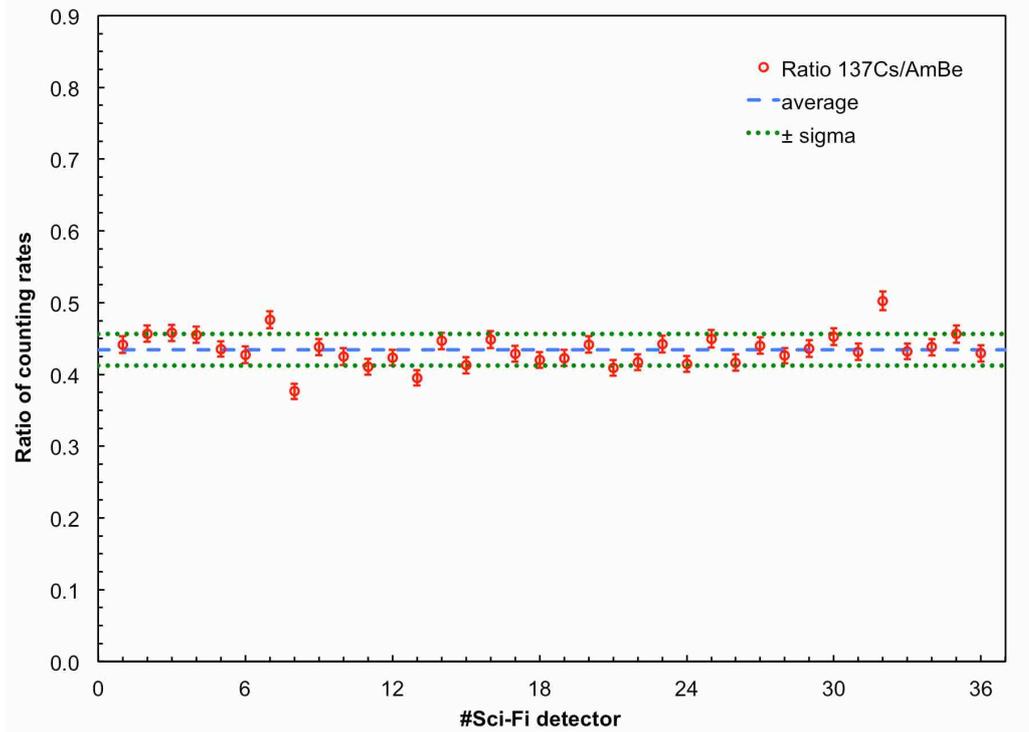

**Figure 17.** Ratio between the observed counting rates respectively with the $^{137}$Cs and the AmBe sources. The results indicate a quite reasonably constant behavior. The error bars represent the statistical uncertainty; the blue dashed line is the average ratio; the green dotted lines indicate ±1 standard deviations from the average.

The spectrum of the gamma rays exiting the source box was measured at several distances by means of a cylindrical Lanthanum Bromide scintillator (LaBr$_3$, in short LaBr), 3.8 cm diameter and 5 cm length, placed at mid-height with respect to the box. LaBr naturally contains about 0.09% of the $^{138}$La isotope, which is unstable and emits gamma rays at 1435 keV, and also impurities of alpha emitters which produce counts around 2-3 MeV equivalent energy. The spectrum measured at 4 cm distance from the box wall is plotted in **Figure 18**, along with the environment background measured far from the source. For all the following evaluations the background spectrum was subtracted from the measurements.

The LaBr spectra were numerically convoluted with the simulated detection efficiency of the SciFi (**Figure 2**) and rescaled for the different active area, in order to estimate the counting rate to be expected on the SciFi detectors at several distances. In **Figure 19** we reported as an example the spectrum of **Figure 18** after background subtraction and folding with the SciFi efficiency. The integral of such a spectrum is the expected count rate on a SciFi at the same distance. The same measurements were done using SciFi number 12, which has an average behavior (**Figure 12**, **Figure 14**, **Figure 16**), and the net count rates (i.e. background subtracted) were reported as a function of the distance in **Figure 20**, along with the corresponding equivalent data inferred from the LaBr measurements.

The data were fitted with the inverse squared distance function

$$f(x) = a \,/\, (x+d)^2 + b, \qquad (9)$$

where a is a scale constant, b is the background and the offset d is the distance between the box wall and the source inside. The fitting curves are also shown in **Figure 20**. Not surprisingly the results, listed in **Table 3**, are mutually consistent as they exhibit a very close scale factor and a null background. As for the offset we observe that the effective position of the source inside the box is not well known (the source cannot be easily han-dled due to safety restrictions because of its huge activity), and both values are realistic. However, such a difference arises from the different response at close distance from the box, due to the relevant shape difference between the two detectors.

A final test was done by measuring the counting rate in 100 s for the SciFi detector n.12 in twenty-two positions, on top and around the AmBe source box as sketched in **Figure** 21. The measured count rates in the twenty-two positions are shown in **Figure** 22.

**Table 3.** Results of an inverse squared distance fit to the SciFi and LaBr count rates as a function of the distance from the source wall.

|       | a (scale constant) | b (background) | d (source position offset) |
|-------|--------------------|----------------|----------------------------|
| SciFi | 9.73               | 0.00           | 0.57 m                     |
| LaBr  | 9.44               | 0.00           | 0.47 m                     |

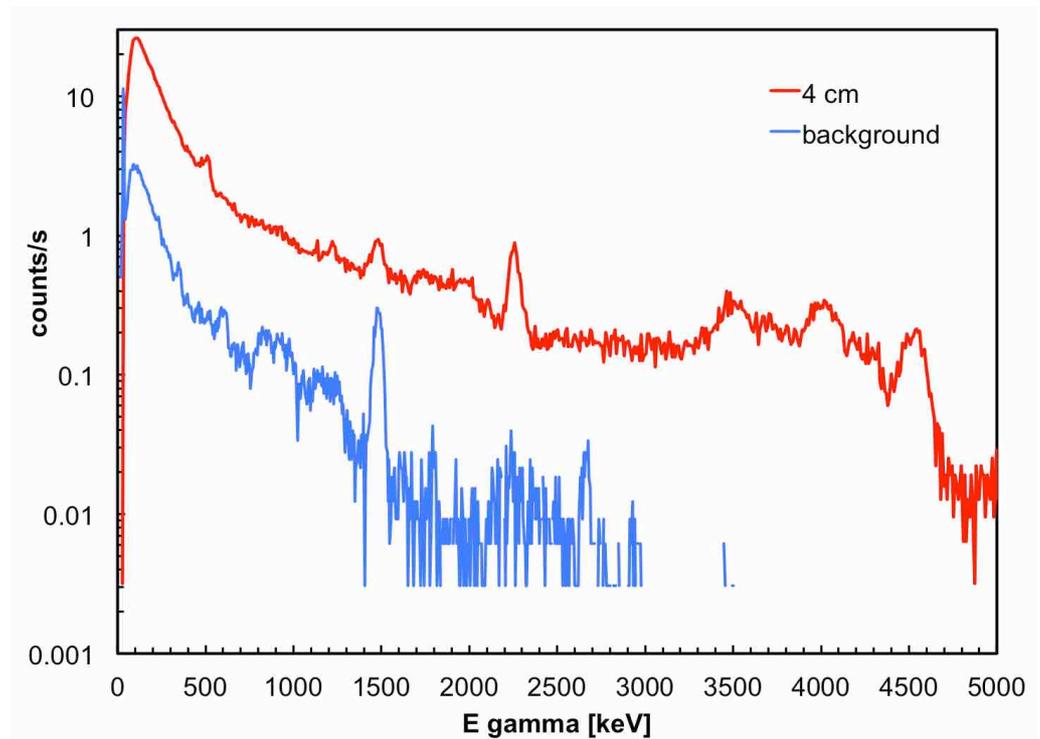

**Figure** 18. Energy spectrum of the gamma rays emerging from the source box at 4 cm distance, and of the environment background.

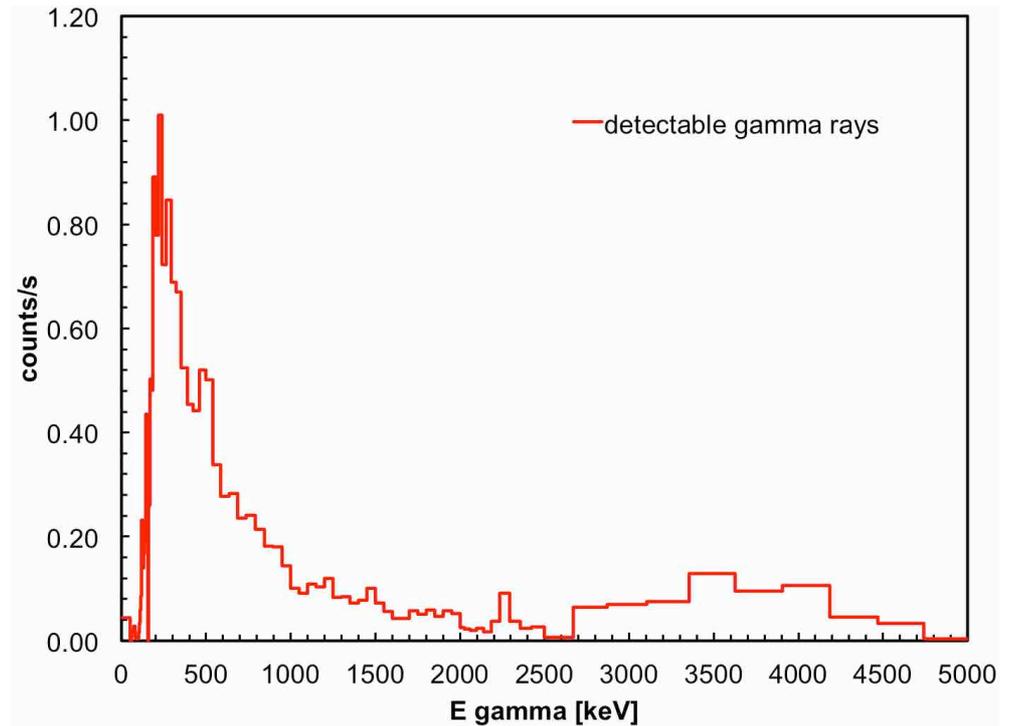

**Figure** 19. The LaBr spectrum at 4 cm of **Figure** 18, after background subtraction and folding with the SciFi efficiency. The integral of this spectrum represents the expected SciFi count rate.

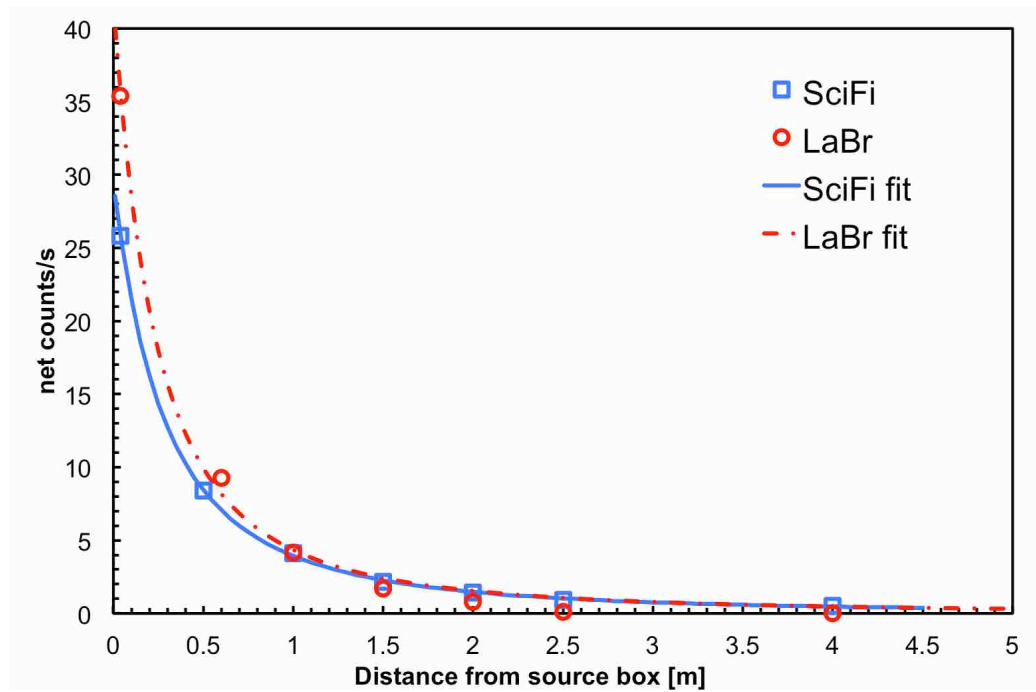

**Figure** 20. Net count rate as a function of the distance from the source wall, for the SciFi n.12 and the LaBr detectors. A data fit is also shown, the fit coefficients are listed in **Table 3**.

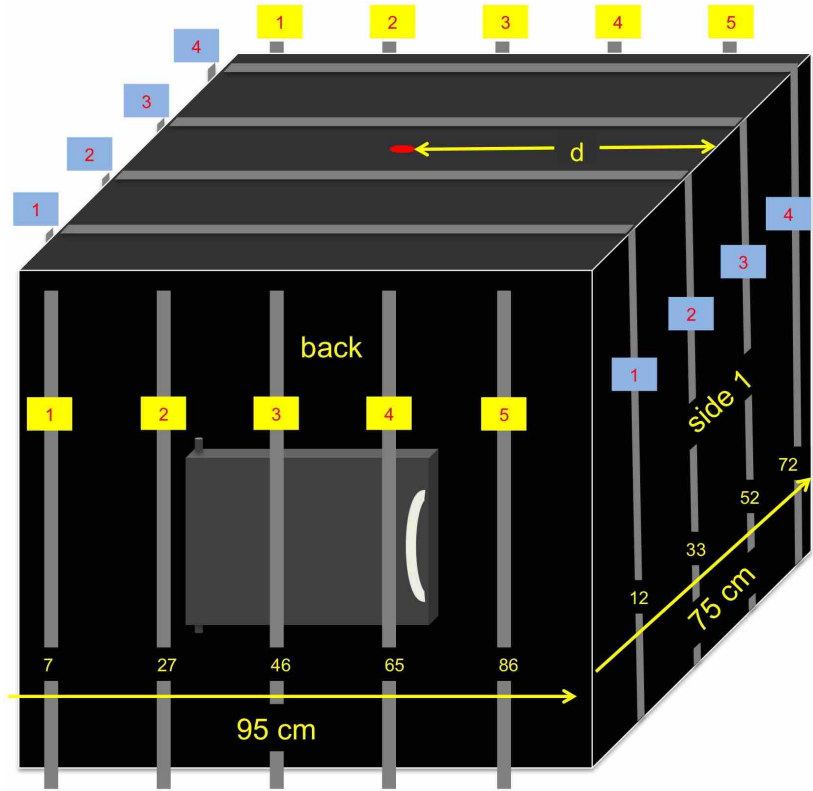

**Figure 21.** Sketch of the twenty-two positions on top and around the AmBe source box where the measurements with the SciFi detector were done. The projection of the source position on the top plane is shown, the exact value of d is not known.

| | 29.7 | 34.3 | 40.2 | 33.1 | 28.7 | |
|---|---|---|---|---|---|---|
| 21.9 | | 21.9 | | | | 20.1 |
| 26.1 | | 31.6 | | | | 29.6 |
| 29.2 | | 31.2 | | | | 29.9 |
| 17.0 | | 19.3 | | | | 19.3 |
| | 13.8 | 22.8 | 29.8 | 23.3 | 16.9 | |

**Figure 22.** Count rates (cps) of the SciFi n.12 in the twenty-two positions on top and around the AmBe source box sketched in **Figure 21**.

## 4. Discussion

In the previous sections we have shown that the SciFi detectors can fruitfully be employed to detect and count gamma rays, in a rather wide energy range. Indeed, by folding the spectra measured by a LaBr detector with the simulated SciFi detection efficiency one obtains count rates in quite a reasonable agreement with the data measured by the SciFi detectors. Apart from very low energy, where the detection efficiency drops below 1%, in the energy range between 150 and 5000 keV it is between 1% and 2%. On the one hand this means that the total number of counts one obtains if assuming an intermediate efficency value of ≈ 1.3% is reasonably correct. On the other hand such a low detection efficiency insures that the detector will not be saturated up to a high gamma flux. The equivalent dose rate in the configuration of **Figure** 11 is of the order of 0.6 μSv/h, with the count rate below 15 cps. A SciFi detector can easily withstand $10^5$-$10^6$ cps, that would roughly correspond to an equivalent dose rate of the order of 4-40 mSv/h. These results prove that SciFi detectors can successfully be employed in the MICADO project in order to monitor radwaste drums in a storage site in the medium/long term.

An interesting exercise has been done in such a framework, to investigate the possibility of monitoring a reference case of radwaste drum. Such a case consists of a Radioactive Waste Package assembly, enclosed in a standard 220 liter drum (86 cm height, 57 cm diameter) with a PVC (poly(vynyl-chloride)) matrix. The drum, sketched in **Figure** 23 along with four SiLiF detectors around it, is supposed to contain $^{60}$Co (55 kBq), $^{133}$Ba (5 kBq), $^{134}$Cs (2 kBq), $^{137}$Cs (2 kBq), $^{152}$Eu (132 kBq), $^{154}$Eu (10 kBq), $^{241}$Am (25 kBq). All of these isotopes are assumed to be dispersed in the matrix, and to give rise to an equivalent dose rate of 2.4 μSv/h at 1 cm [24].

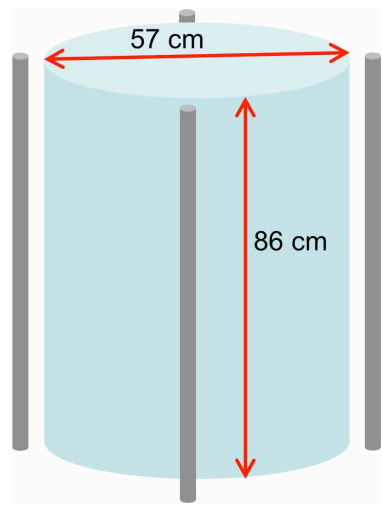

**Figure** 23. Sketch of the simulated radwaste package with four SciFi detectors around it.

We observe that the majority of the dose rate is due to $^{60}$Co and $^{152}$Eu, which have much higher activity and emit gamma rays from 344 to 1330 keV. Assuming the same detection efficiency as in the case of $^{22}$Na (Eq.2 and **Figure** 10), the cps to equivalent dose rate coefficient is ≈ 1.18 (μSv/h) / cps, and vice versa the equivalent dose rate to cps coefficient is ≈ 0.85 cps / (μSv/h) [23]. This implies that the expected net count rate in a SciFi is of the order of 2.4 x 0.85 ≈ 2 cps, which has to be summed to the background count rate which should be of the same order of magnitude.

In one minute the statistical uncertainty of the measured counting rate would be better than 10%, and in ten minutes better than 3%. This means that any initial counting asymmetry between the four fibers is quickly appreciable, as well as any change that might occur due to internal displacement of the waste and/or deterioration of the drum. As for the radiation damage, we assume it is neglibible as according to the test results reported in [25] the SciFi detectors would not be affected even after a hundred years of exposure at their highest dose rate counting. Possible failures in the long term will most likely be due to electronics.

## 5. Conclusions

The simulations, tests and measurements we have done allowed us to show that the SciFi technology is a good candidate for gamma radiation monitoring of radioactive waste. The 36 detectors we built have a reasonably uniform behavior, in light of their robustness, low cost, and simple construction based on commercial components. We are now planning to test them soon in a real radwaste storage site in the framework of the MICADO project.

**Author Contributions:** conceptualization and supervision P.F.; detector design P.F., F.L., A.P., L.C.; detector construction F.L., A.P. and G.P.; test and measurements F.L., A.P., M.G., L.C. and P.F.; simulations S.L.M.; manuscript preparation P.F. All authors have read and agreed to the published version of the manuscript.

**Funding:** This work was funded within the framework of European Union's Horizon 2020 research and innovation programme under grant agreement No 847641, project MICADO (Measurement and Instrumentation for Cleaning And Decommissioning Operations). This work was also supported by the Centro Siciliano di Fisica Nucleare e di Struttura della Materia [Grant 02/2019].

**Acknowledgments:** We are grateful to Dr. Marco Ripani for the constant encouragement and for the initial support within the INFN-Energy program that allowed us to start the MICADO project.

**Conflicts of Interest:** The authors declare no conflict of interest.